\documentclass[english,nofootinbib,prd]{revtex4}
\usepackage[latin9]{inputenc}
\usepackage{amsmath}
\usepackage{amssymb}
\usepackage{graphicx}

\makeatletter
\@ifundefined{textcolor}{}
{%
 \definecolor{BLACK}{gray}{0}
 \definecolor{WHITE}{gray}{1}
 \definecolor{RED}{rgb}{1,0,0}
 \definecolor{GREEN}{rgb}{0,1,0}
 \definecolor{BLUE}{rgb}{0,0,1}
 \definecolor{CYAN}{cmyk}{1,0,0,0}
 \definecolor{MAGENTA}{cmyk}{0,1,0,0}
 \definecolor{YELLOW}{cmyk}{0,0,1,0}
}

\usepackage{amsfonts}
\providecommand{\U}[1]{\protect\rule{.1in}{.1in}}

\usepackage{babel}

\makeatother

\usepackage{babel}
\begin{document}

\title{On the invariance of the relative rest in doubly special relativity}

\author{Gianluca Mandanici}

\email{Gianluca.Mandanici@unibg.it}

\affiliation{Università degli Studi di Bergamo, Facoltà di Ingegneria, Viale Marconi
5, 24044 Dalmine (Bergamo) Italy.}
\begin{abstract}
In the framework of the most-studied doubly special relativity models
the use of the naive formula $v=dE/dp$ has been argued to lead to
inconsistencies connected to different rules of transformation, under
boosts, of particles with the same energy but with different masses.
In this paper we show that, at least in 1+1 dimensions, doubly special
relativity can be formulated in such a way that the formula $v=dE/dp$
is fully consistent with the invariance of the relative rest, easily
fitting to the relativity principle. It is also argued that, always
in 1+1d, is not necessary to renounce to the usual (commutative) Minkowski
space-time endowed with energy-independent boost transformations.
The compatibility of the approach with superluminal propagation, with
linear addition rule for energy, and possible extensions to 3+1 dimensions
are also discussed. 
\end{abstract}
\maketitle

\section{Introduction}

\label{S1}

In the quest for a quantum theory of gravity notable interest has
been attracted over the last decade and more, from scenarios in which
the Lorentz/Poincaré symmetry is deformed (but not broken) at the
Planck scale. A much studied class of these scenarios is often referred
in literature as doubly special relativity$\,$\cite{AmelinoCamelia:2000ge,AmelinoCamelia:2000mn,Magueijo:2001cr,KowalskiGlikman:2002we,AmelinoCamelia:2003ex}
and typically consists in a two-scale extension of Einstein's special
relativity, being the second invariant scale eventually connected
with the Planck energy/length. At first, doubly special relativity
has been formulated in the energy-momentum sector, however, over the
years, many attempts have been made to consistently extend the theory
from the energy-momentum sector to the space-time sector (for a recent
review see \cite{AmelinoCamelia:2010pd}). A number of these attempts
are based on the assumption of a underling non-commutative space-time
as a consequence of the fuzziness of space-time at length scale of
the order of the Planck length (see e.g.$\,$\cite{Lukierski199590,KowalskiGlikman:2002jr,Mignemi:2003dm,Galan:2007ns,Ghosh:2007ai}).
Other attempts instead assume a underlying commutative space-time
(see e.g.$\,$\cite{Magueijo:2002am,Kimberly:2003hp}). What is common
within most of these approaches is that one obtains energy-dependent
space-time transformations with the consequence that particles with
the same speed, but with different masses, acquire relative motion
when considered from different inertial observers. This unclassical
feature of space-time transformations could appear in contrast with
the relativity principle especially for the apparent loss of locality$\,$\cite{Hossenfelder:2010tm},
that is recovered only in a form that depends on the choice of the
inertial frame$\,$\cite{AmelinoCamelia:2011bm,AmelinoCamelia:2010qv,AmelinoCamelia:2011gy}.
A central role in the delicate transition from the energy-momentum
sector to the space-time sector of doubly special relativity is played
by the formula that expresses the speed of a free particle as a function
of its energy and its momentum. While the first proposals$\,$\cite{AmelinoCamelia:2000ge,AmelinoCamelia:2000mn,Magueijo:2001cr}
have assumed the validity of the usual formula $v=dE/dp,$ in$\,$\cite{Kosinski:2002gu}
this formula been argued to be incompatible with the standard hamiltonian
formalism. In$\,$\cite{Mignemi:2003ab} starting from the request
that the formula of the particle speed should not depend on the mass
of the particles, the author was led to rule out the formula $v=dE/dp.$
Similar results have been obtained in$\,$\cite{Aloisio:2004rd} where
has also been claimed that the dependence of the velocity on the mass
of the boosted particle appear unavoidable using the mentioned formula.
A study of velocity of particle in non-commutative space-time still
compatible with the formula $v=dE/dp$ is instead that of$\,$\cite{AmelinoCamelia:2002tc}.
The main aim of this paper is just to address the issue of the relative
rest in doubly special relativity and its compatibility with the standard
formula $v=dE/dp.$ The structure of the paper is the following. In
section$\,$\ref{S2} we discuss the implications of the request of
covariance of the velocity in the form $dE/dp$ for the energy-momentum
sector of the most-studied doubly special relativity models, providing
explicit examples of covariance. In section$\,$\ref{S3} we propose
a construction of the space-time sector from given energy-momentum
sectors. In section$\,$\ref{S9} we address the issues of the compatibility
of the approach with deformations of the boost action on the particle
speed, with linear addition rule for energy and we also discuss a
possible extension to 3+1 space-time dimensions. Finally, in section$\,$\ref{S10},
we present our conclusions.

\bigskip{}

\section{Momentum space formulation}

\label{S2}

\subsection{The classical relativistic picture of relative rest}

The first working assumption of this paper is that \textit{particles
that do not possess relative motion in a given reference frame do
not acquire relative motion in any other (inertial) reference frame}.
In the usual formulation, for infinitesimal transformations, the speed
of a boosted particle $v^{\prime}$ is obtained from the speed of
the particle $v$ by means of the formula 
\begin{equation}
v^{\prime}(\xi)\simeq v+\xi\left\{ N,v\right\} ,\label{speedtransformationrule}
\end{equation}
where $\xi$ is the boosting parameter and $\left\{ N,v\right\} $
is the infinitesimal boost action. In this notation, in order to obtain
the invariance of the relative rest\textit{, }the boost operator must
be function only of the particle speed $v$ and of (possible) fundamental
constants $c_{1},...,c_{n}$ of the model 
\begin{equation}
\left\{ N,v\right\} =K(v,c_{1},...,c_{n}).\label{boostv}
\end{equation}

Notice that this request is satisfied in Galilei relativity where
no invariant velocity scale is present. In fact in Galilei relativity,
being $E=p^{2}/(2m)$, $v=dE/dp=p/m$ and $\left\{ N,p\right\} =m$,
one finds 
\begin{equation}
\left\{ N,v\right\} =\left\{ N,p\right\} \frac{dv}{dp}=1,
\end{equation}
so that $K(v)=1.$ Also in Einstein's special relativity, where the
invariant velocity scale $c$ is present, being $E^{2}=c^{2}p^{2}+c^{4}m^{2}$,
$v=dE/dp=c^{2}p/E$ and $\left\{ N,p\right\} =E/c^{2},$ one finds
\begin{equation}
\left\{ N,v\right\} =\left\{ N,p\right\} \frac{dv}{dp}=1-c^{2}p^{2}/E^{2},
\end{equation}
i.e. $K(v,c)=1-v^{2}/c^{2}.$ Again the function $K(v)$ depends only
on the particle speed and on the invariant scale $c$. In this latter
case one also gets $K(c,c)=0$ consistently with the fact that $c$
is not only an invariant parameter of the theory but, rather, it is
the invariant speed of (massless) particles. In both these known cases
$K(v)$ depends only on the particle speed, so that two particles
with the same speed in a given reference frame $v_{1}=v_{2}=v$ do
not acquire relative motion in any other inertial reference frame
\begin{equation}
v_{2}^{\prime}-v_{1}^{\prime}\simeq\xi\lbrack K(v,c)-K(v,c)]=0,
\end{equation}
independently on their masses.

\subsection{Relative rest in the most studied doubly special relativity models}

In most approaches to Planck-scale relativity there are strong departure
from the classical picture as outlined above. In fact if one adopts
the formula%
\footnote{We emphasize here that this formula that works well in classical physics
and quantum mechanics can also be derived in any Hamiltonian approach
$v=\dot{x}=$ $\left\{ x,E(p)\right\} $ also compatible with $\left\{ x,p\right\} =1$,
being $v=\left\{ x,E\right\} =\left\{ x,p\right\} dE/dp=dE/dp.$ %
}
\begin{equation}
v=\frac{dE}{dp},\label{speed}
\end{equation}
one finds that in general $K=K(v,c,\lambda cm,\lambda p)$, where
$\lambda$ is the Planck-scale parameter, eventually identified with
the inverse of the Planck momentum. This means that the action of
the boosts on the speed of a particle depends not only on the particle
speed itself but also on the particle mass. The main consequence of
this dependency is that particles having the same speeds $\overline{v}$
in a given reference frame could acquire relative motion in a different
reference frame, if they have different masses. In fact, being in
general $K(\overline{v},c,\lambda cm_{1},\lambda p_{1})\neq K(\overline{v},c,\lambda cm_{2},\lambda p_{2}),$
one finds 
\begin{equation}
\overline{v}_{1}^{\prime}-\overline{v}_{2}^{\prime}=\xi\lbrack K(\overline{v},c,\lambda cm_{1},\lambda p_{1})-K(\overline{v},c,\lambda cm_{2},\lambda p_{2})]\neq0.
\end{equation}

This effect has been analyzed in$\,$\cite{Aloisio:2004rd} in the
framework of the so-called DSR1 model proposed in$\,$\cite{AmelinoCamelia:2000ge,AmelinoCamelia:2000mn}
that admits energy-momentum dispersion relation of the type 
\begin{equation}
E=c\left|\lambda\right|^{-1}\ln\left[\frac{\cosh\left(\lambda mc\right)+\sqrt{\cosh^{2}\left(\lambda mc\right)-(1-\lambda^{2}p^{2})}}{(1-\lambda^{2}p^{2})}\right],\label{DSR1}
\end{equation}
but the same effect is also present in the other most-studied doubly
special relativity model, the so-called DSR2 model, proposed in$\,$\cite{Magueijo:2001cr}.
Again can be easily shown that from energy-momentum dispersion relation
of the form 
\begin{equation}
\frac{E^{2}-p^{2}c^{2}}{(1-\lambda E/c)^{2}}=\frac{m^{2}c^{4}}{(1-\lambda mc)^{2}},\label{MSDDDR}
\end{equation}
using (\ref{boostv}) and (\ref{speed})\ one finds 
\begin{equation}
K(v,c,\lambda m)=\frac{1-v^{2}/c^{2}}{1+\lambda cm/\sqrt{(1-\lambda cm)^{2}-v^{2}(1-2\lambda cm)/c^{2}}}.\label{KDSR2}
\end{equation}

From eq.(\ref{KDSR2}) follows that two particles at rest in a given
reference frame ($v_{1}=v_{2}=0$), will acquire a relative motion:
\begin{equation}
v_{1}^{\prime}-v_{2}^{\prime}=\xi\lbrack K(0,c,\lambda m_{1})-K(0,c,\lambda m_{2})]\simeq\xi\lambda c(m_{1}-m_{2})\neq0.\label{relmotion}
\end{equation}

The dependency of the boosts on the mass of the particle does not
necessarily imply a violation of the relativity principle but, at
least, suggests that locality should become a concept relative to
the particular choice of the reference frame (see e.g.$\,$\cite{AmelinoCamelia:2011bm,AmelinoCamelia:2010qv,AmelinoCamelia:2011gy}).
In the rest of this paper we wont address the point of whether or
not effects of the type implied by formula (\ref{relmotion}) can
be incorporated in a relativistic scheme. Rather we intend to focus
here on what could be the features of doubly special relativity models
in which effects implied by formula (\ref{relmotion}) are absent
at all, like they are in the known classical cases.

\subsection{Approaching relative rest invariance in doubly special relativity}

\subsubsection{General theoretical framework}

Once we have discussed the way in which the problem manifests in doubly
special relativity we can move forward looking for a solution. We
start requiring the covariance under boosts of the energy-momentum
dispersion relation $E=E(p).$ From this request it follows that it
must be 
\begin{equation}
\left\{ N,E\right\} =\left\{ N,p\right\} \dfrac{dE}{dp},\label{CovRelDisp}
\end{equation}
that fixes the relation between the action of the boosts on the energy
and the action of the boosts on the momentum. Our next request is
that $\left\{ N,v\right\} $ be a function of only the particle speed,
expressed as in eq.(\ref{speed}), from which one gets the following
formula 
\begin{equation}
\left\{ N,v\right\} =\left\{ N,p\right\} \dfrac{d^{2}E}{dp^{2}}=K\left(\frac{dE}{dp}\right).\label{eq:RR}
\end{equation}

The validity of eq.(\ref{eq:RR}) directly implies that particles
with the same speed (i.e. at rest in a given reference frame) wont
acquire any relative motion in any other reference frame. Thus the
request that the boosted velocity depends only on the particle speed
fixes the infinitesimal action of the boost on the momentum in the
form: 
\begin{equation}
\left\{ N,p\right\} =K\left(\frac{dE}{dp}\right)/\dfrac{d^{2}E}{dp^{2}}.\label{CovSpeedExpress}
\end{equation}

Finally substituting (\ref{CovSpeedExpress}) in (\ref{CovRelDisp})
one obtains the infinitesimal action of the boost on the particle
energy 
\begin{equation}
\left\{ N,E\right\} =K\left(\frac{dE}{dp}\right)\dfrac{dE}{dp}/\dfrac{d^{2}E}{dp^{2}}.\label{BoostEnergyAction}
\end{equation}
Formulas (\ref{CovSpeedExpress})-(\ref{BoostEnergyAction}) together
with (\ref{speed}) are the main working tools of the analysis presented
in the rest of the paper.

\subsubsection{A first-order example in the Planck scale}

We are now ready to apply the procedure outlined above to a general
Planck-scale deformed dispersion relation. As a warm up we start the
analysis with a first order deformation. The dispersion relation that
we consider here is of the type of those studied in\ \cite{AmelinoCamelia:1997gz,Gambini:1998it},
that is a most studied energy-momentum dispersion relation: 
\begin{equation}
E^{2}-c^{2}p^{2}+\lambda p^{3}c^{2}=m^{2}c^{4}.\label{fodr}
\end{equation}

\begin{figure}[ht]
\includegraphics[scale=0.6]{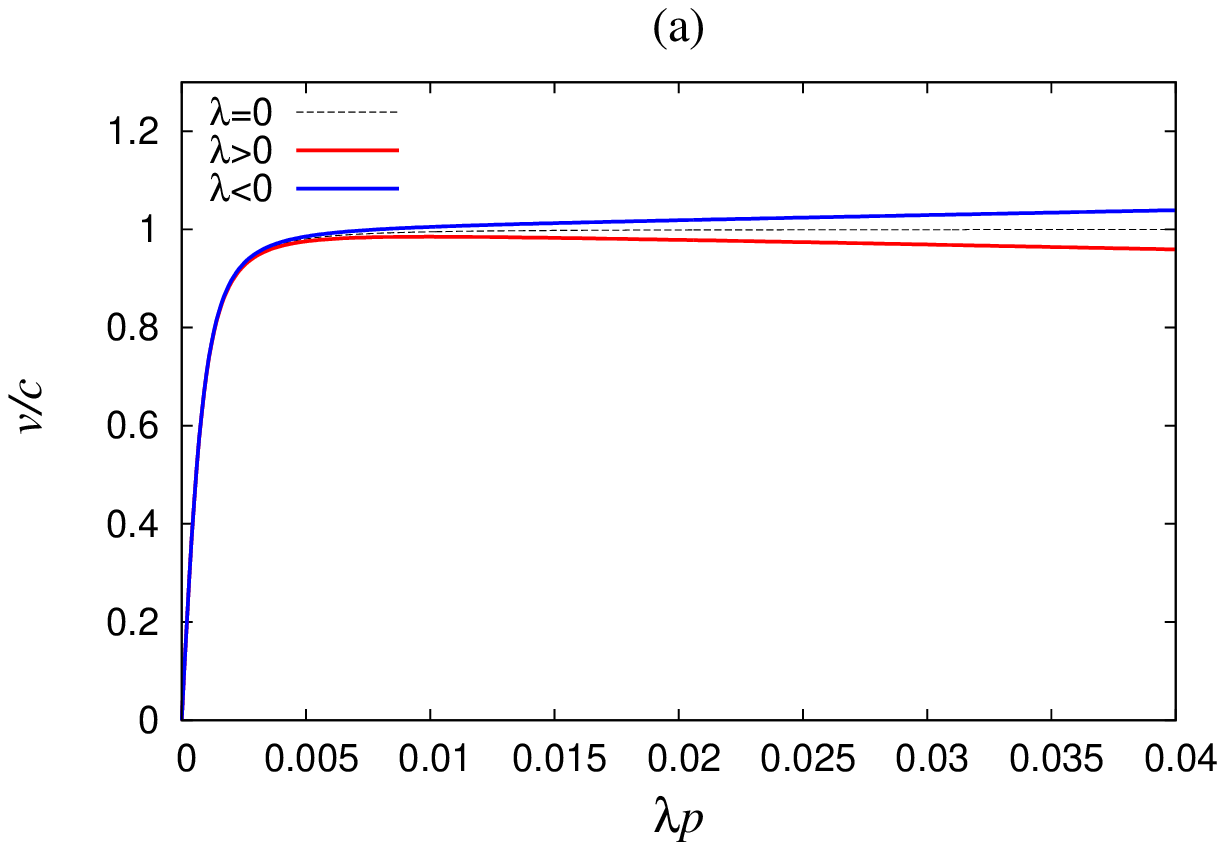}\includegraphics[scale=0.6]{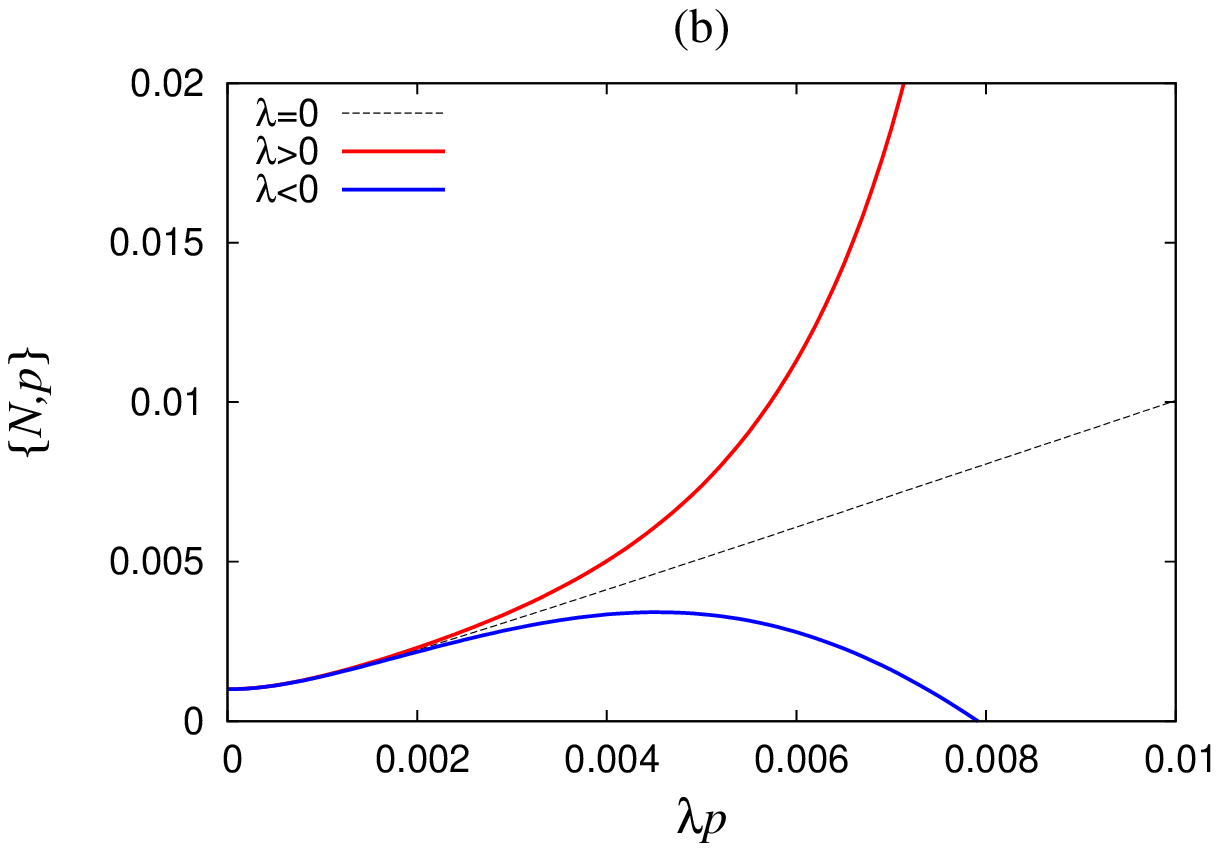}
\caption{(a) $v/c$ vs $\lambda p$ and (b) $\left\{ N,p\right\} $ vs $\lambda p$
for the dispersion relation of eq.(\ref{fodr}) with $mc\lambda=0.001.$
The black-dashed lines ($\lambda=0$) are the predictions of special
relativity. }

\label{Fig1} 
\end{figure}

Particle speed can be calculated from the dispersion relation using
eq.(\ref{speed}) obtaining 
\begin{equation}
v=\dfrac{dE}{dp}=\frac{c^{2}p(1-3/2\lambda p)}{\sqrt{c^{4}m^{2}+p^{2}c^{2}-\lambda c^{2}p^{3}}},
\end{equation}
that, for massless particles, assumes the form $v\simeq c(1-\lambda p)$,
at the first order in $\lambda.$ To derive the action of the boosts
on the energy and on the momentum we use formulas (\ref{CovSpeedExpress})-(\ref{BoostEnergyAction}).
The form of $K(v)$ is in principle arbitrary unless it provides the
correct Minkowski limit. However, in this section, we will assume
an undeformed expression for $K(v)$. Thus fixing $K(v)=1-v^{2}/c^{2}$,
one finds for the action of the boost: 
\begin{align}
\left\{ N,p\right\}  & =\frac{\left[4c^{2}m^{2}+p^{3}\lambda\left(8-9p\lambda\right)\right]\sqrt{c^{2}m^{2}+p^{2}-\lambda p^{3}}}{cp^{3}\lambda(-4+3p\lambda)+4c^{3}m^{2}\left(1-3\lambda p\right)},\label{Np}\\
\left\{ N,E\right\}  & =\left\{ N,p\right\} \frac{c^{2}p(1-3/2\lambda p)}{\sqrt{c^{4}m^{2}+p^{2}c^{2}-\lambda c^{2}p^{3}}},\label{NE}
\end{align}
whose behavior, as a function of the particle momentum $p$, is reported
in Fig.$\,$\ref{Fig1}.

It is easy to check that $i)$ the dispersion relation (\ref{fodr})
is invariant under boost action and that $ii)$ we have $v_{1}^{\prime}-v_{2}^{\prime}\simeq\xi\left[K(0,c,\lambda cm_{1})-K(0,c,\lambda cm_{2})\right]=0$
for all the possible values of the masses $m_{1}$ and $m_{2}$ of
the particles involved. Thus the state of relative quite is preserved
for all the particles in all the inertial frames. The entire analysis
here reported is performed in the energy-momentum space. The only
connection with space-time is provided by formula (\ref{speed}) that
defines the particle speed. In the case of $\lambda=0$ one recovers
the usual special relativistic formulas. The case $\lambda<0$, that
in principle should correspond to superluminal propagation (i.e. to
$v>c$), in our approach returns $\left\{ N,p\right\} \simeq\left\{ N,E\right\} \simeq0$
as soon as $p\simeq p_{I}=\sqrt[3]{m^{2}c^{3}\lambda^{-1}/2}$, as
it is clear both from the analytical expressions (\ref{Np})-(\ref{NE})
and from Fig.$\,$\ref{Fig1}. This means that $p_{I}$ is an invariant
momentum ($p_{I}^{\prime}\simeq p_{I}$), and the corresponding energy
an invariant energy ($E_{I}^{\prime}\simeq E_{I}$). The reasons why
one gets invariant energy and momentum from eqs.$\,$(\ref{Np})-(\ref{NE})
is that as soon as the momentum approaches $p_{I}$ the speed of the
particle involved approaches the invariant speed $c,$ being in fact
$\left\{ N,v(p_{I})\right\} \simeq0$. Notice that the invariant momentum
$p_{I}$ depends on the mass of the particle involved and that it
is different from the Planck momentum, as far as the mass of the particle
involved is different from the Planck mass. The case $\lambda>0$
corresponds to subluminal propagation (i.e. to $v<c$). In this case
the particle never approaches $v\simeq c$ and thus there is no invariant
momentum. However there is a maximum momentum since, as soon as $p\simeq p_{max}=\sqrt[3]{m^{2}c^{3}\lambda^{-1}}$
, one gets $\left\{ N,p_{max}\right\} \simeq\left\{ N,E(p_{max})\right\} \simeq\infty$.

\subsubsection{Two all-order examples: the Magueijo-Smolin dispersion relation and
the $\kappa$-Poincaré dispersion relation.}

We are now ready to apply our procedure to all-order (in the Planck
scale) dispersion relations. The first type of dispersion relation
that we consider is the Magueijo-Smolin dispersion relation (\ref{MSDDDR}).
Again using formula (\ref{speed}) one gets the particle speed: 
\begin{equation}
v=\dfrac{dE}{dp}=\frac{c^{2}p(1-mc\lambda)^{2}}{\left(1-2\lambda cm\right)E+m^{2}c^{3}\lambda}.
\end{equation}

\begin{figure}[ht]
\includegraphics[scale=0.6]{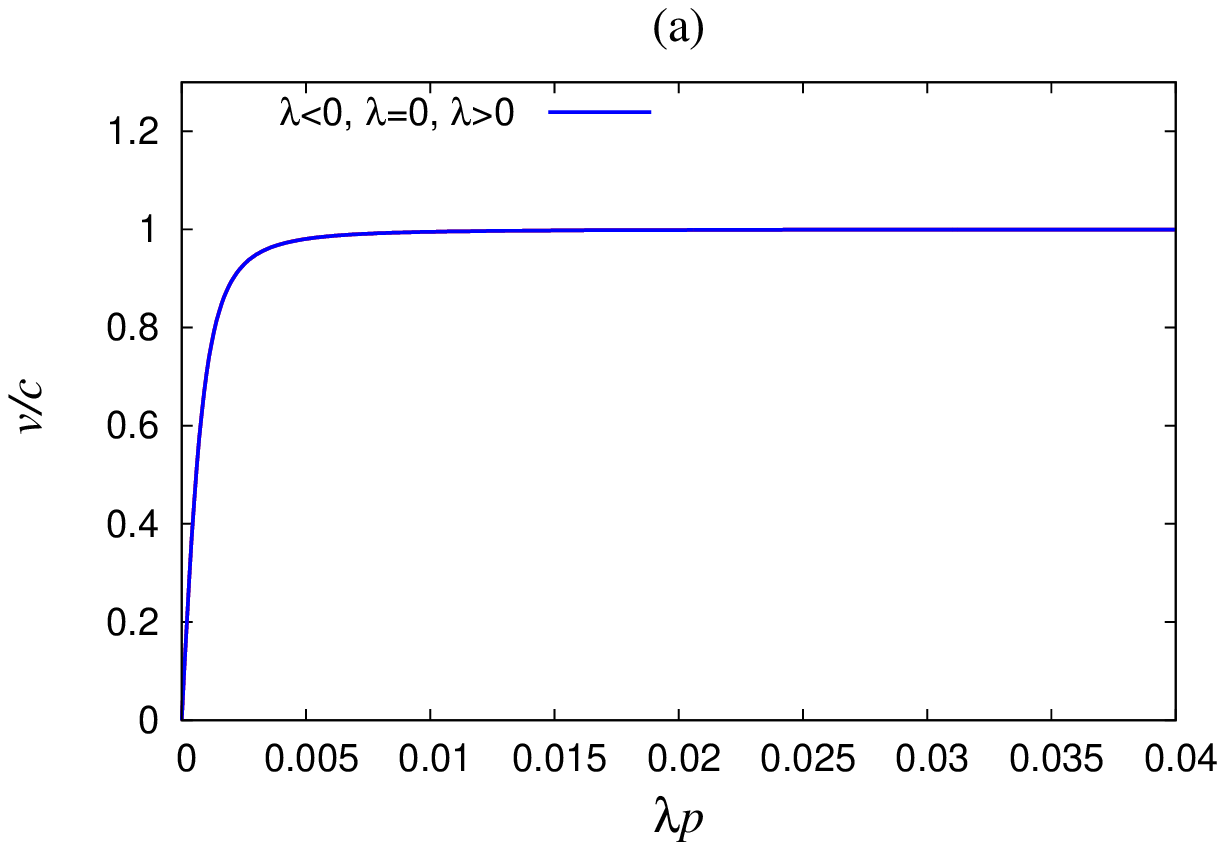} \includegraphics[scale=0.6]{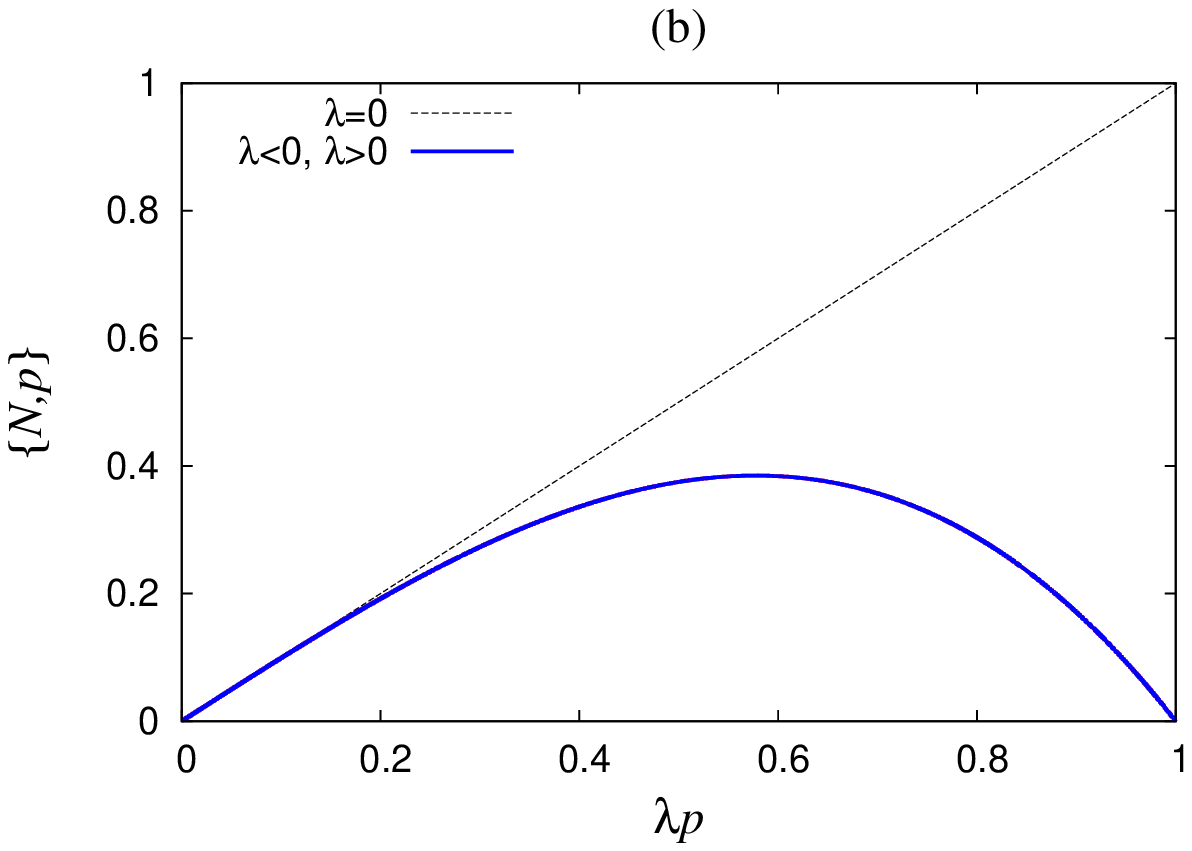}
\caption{(a) $v/c$ vs $\lambda p$ and (b) $\left\{ N,p\right\} $ vs $\lambda p$
for the dispersion relation of eq.(\ref{MSDDDR}) with $mc\lambda=0.001.$
The black-dashed line ($\lambda=0$) is the prediction of special
relativity.}

\label{Figura2} 
\end{figure}

From eqs.(\ref{CovSpeedExpress})-(\ref{BoostEnergyAction}) we obtain
that the transformation rules between inertial observers are given
by the actions 
\begin{align}
\left\{ N,p\right\}  & =\frac{E(1-2mc\lambda)+m^{2}c^{3}\lambda}{c^{2}(1-mc\lambda)^{2}}(1-\lambda^{2}p^{2}),\\
\left\{ N,E\right\}  & =p(1-\lambda^{2}p^{2}),
\end{align}
whose behavior is reported in Fig.$\,$\ref{Figura2}. \bigskip{}

As expected as $v$ approaches the invariant speed $c$, the energy
and the momentum approaches the invariant energy $c\lambda^{-1}$
and the invariant momentum $\lambda^{-1}$ respectively, as in the
case of the original boost actions discussed in\ \cite{Magueijo:2001cr}
but now, differently from\ \cite{Magueijo:2001cr}, it is easy to
check that also 
\begin{equation}
v_{1}=v_{2}\Rightarrow v_{1}^{\prime}=v_{2}^{\prime},
\end{equation}
independently on the particle masses. Notice that it is also possible
to express the energy and the momentum in terms of the mass and the
speed of the particle 
\begin{align}
p\left(v\right) & =\frac{mv}{\sqrt{\left(1-\lambda mc\right)^{2}-v^{2}(1-2mc\lambda)/c^{2}}},\label{eq:pv_ev1}\\
E(v) & =\frac{1}{1-2\lambda mc}\dfrac{m\left(1-\lambda cm\right)^{2}c^{2}}{\sqrt{\left(1-\lambda mc\right)^{2}-v^{2}(1-2mc\lambda)/c^{2}}}-\frac{m^{2}c^{3}\lambda}{1-2\lambda cm}.\label{eq:pv_ev2}
\end{align}

\begin{figure}[ht]
\includegraphics[scale=0.6]{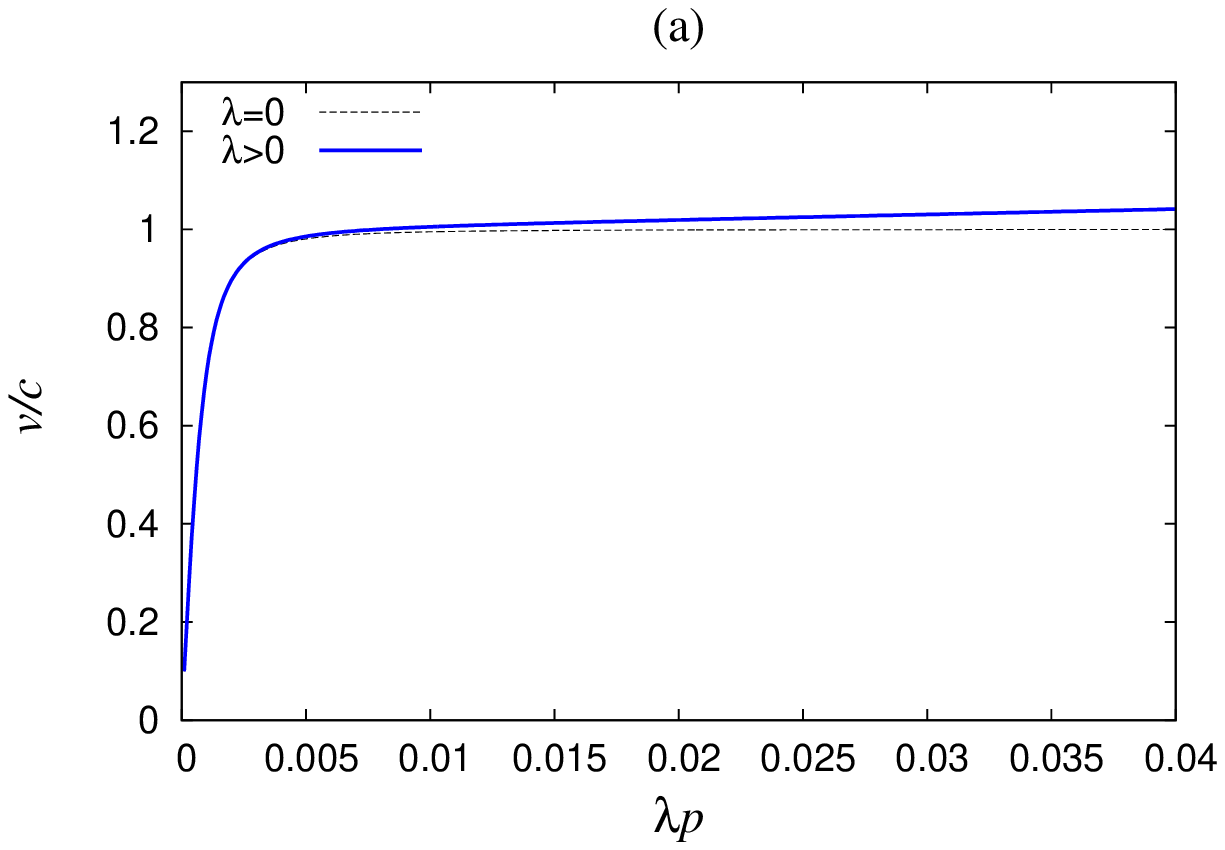} \includegraphics[scale=0.6]{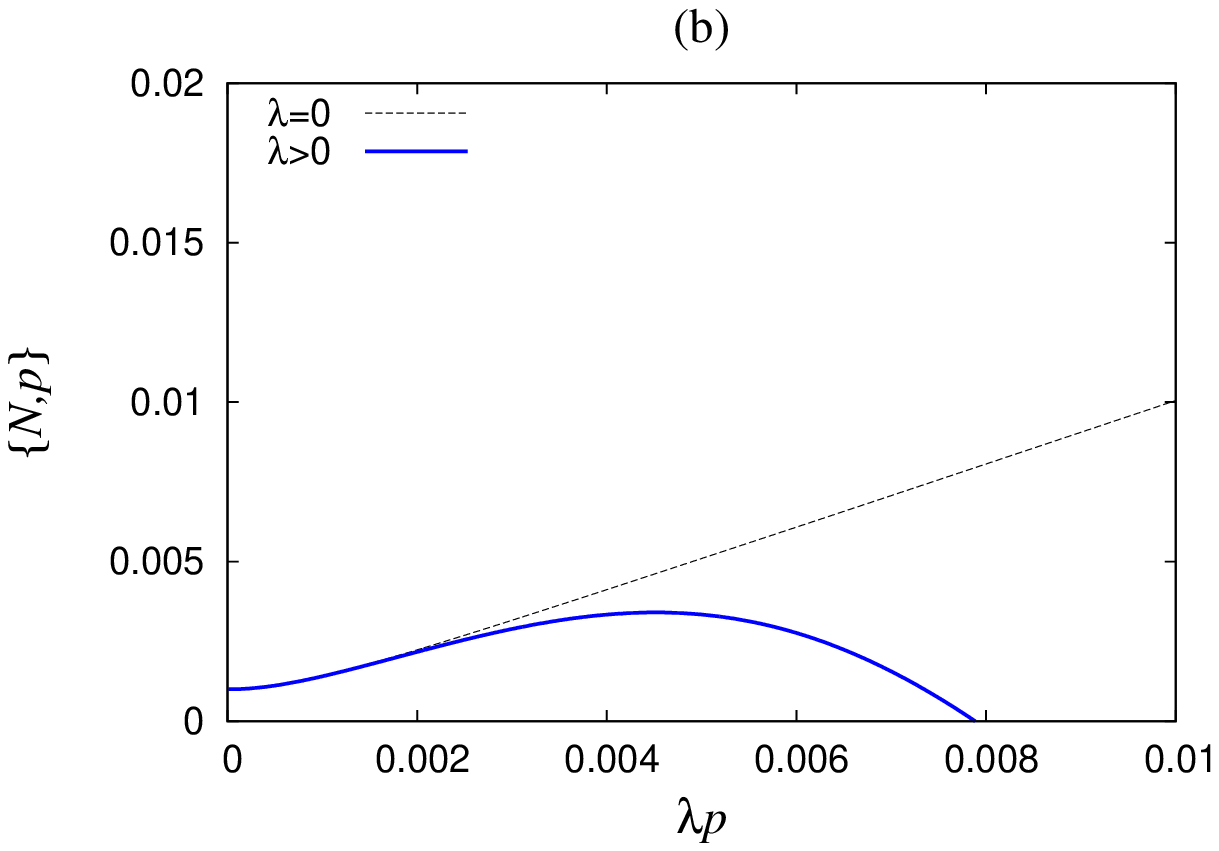}
\caption{(a) $v/c$ vs $\lambda p$ and (b) $\left\{ N,p\right\} $ vs $\lambda p$
for the dispersion relation of eq.(\ref{DSR1}) with $mc\lambda=0.001.$
The black-dashed lines ($\lambda=0$) are the predictions of special
relativity. }

\label{Figura3} 
\end{figure}

The second example of all-order dispersion relation that we want to
consider here is the $\kappa$-Poincaré motivated dispersion relation
of eq.(\ref{DSR1}). Following the same procedure we can easily find
$v,$ $\left\{ N,p\right\} $ and $\left\{ N,E\right\} $ (again involving
eqs.(\ref{speed}), (\ref{CovSpeedExpress}) and (\ref{BoostEnergyAction})).
Being the analytical expressions rather involved we do not report
them. Instead we represent graphically the results in Fig.$\,$\ref{Figura3}
only for case $\lambda>0,$ since the case $\lambda<0$ does not possess,
in our approach, the proper classical limit (see also the discussion
in\ \cite{AmelinoCamelia:2010pd}).

\section{The space-time sector}

\label{S3}

In the preceding sections we have analyzed the invariance of the relative
rest in the energy-momentum sector. In this section we try to address
the same issue in the space-time sector. The link between the energy-momentum
sector and the space-time sector is played by the definition of the
particle speed given that, according to our assumptions, the following
formula must hold 
\begin{equation}
v=\dfrac{dE}{dp}=\dfrac{dx}{dt}.\label{spacetimespeed}
\end{equation}

The transformation rules in the energy-momentum sector have already
been obtained therefore it is necessary that space-time transforms
accordingly. We request that, as usual, different inertial systems
are connected by uniform motion. Since we are looking for space-time
boost transformations of the form 
\begin{equation}
x'(\xi)\simeq x+\xi\left\{ N,x\right\} ,
\end{equation}
in order to have inertial systems moving at a uniform relative motion
it must be 
\begin{equation}
\left\{ N,x\right\} =t.
\end{equation}

Then the covariance in the energy-momentum sector, together with
the request of covariance of (\ref{spacetimespeed}), fixes the transformation
rule 
\begin{equation}
\left\{ N,dt\right\} =\left\{ N,\dfrac{dx}{v}\right\} =\dfrac{1-K(v)}{v}dt.\label{timeboost}
\end{equation}

Notice that the transformation of the time under boost does not depend
on the mass of the particles involved, it depends only on their speeds.
Notice also that the action of boost on time can be a non-linear function
of the particle speed $v$, depending on the particular form of $K(v)$.
Since we do not want to introduce new invariant velocity scales, we
will assume that $K$ has not to be modified with respect to the standard
special relativistic case.%
\footnote{The possibility of introducing a second velocity scale in this framework
is not void of interest and can deserve further investigations.%
} Adopting the usual form $K(v)=1-c^{-2}v^{2}$ and substituting in
(\ref{timeboost}) we get the action 
\begin{equation}
\left\{ N,t\right\} =c^{-2}x,
\end{equation}
that is the usual infinitesimal action in the Minkowski space-time.
It follows that all the elements regarding the space-time sector are
not affected by the introduction of the Planck scale-parameter. In
particular we get the usual dependence of the boost parameter $\xi$
on the relative velocity $v_{S}$ of the inertial systems, and the
usual velocity composition rule 
\begin{equation}
v_{S}=c\tanh\left(\frac{\xi}{c}\right),\hspace{2cm}v^{\prime}=\frac{v+v_{S}}{1+\dfrac{vv_{S}}{c^{2}}}.
\end{equation}

Notice however that being in general $E(p)\neq E_{SR}(p)$ it follows
that $p=\left(dE/dp\right)^{-1}(v)\neq\left(dE_{SR}/dp\right)^{-1}(v)$
so that the momentum and the energy depend on the speed of the particle
in a deformed way with respect to the special relativistic case, as
already evident from eqs.(\ref{eq:pv_ev1})-(\ref{eq:pv_ev2}).

\section{Further issues}

\label{S9}

In this section we consider other issues related to the construction
developed. The issues that we want to address are \textit{i)} the
possibility of adopting different forms of the function $K(v)$, \textit{ii)}
the compatibility of the formalism with a linear addition law for
energies, \textit{iii)} the possibility to extend the formalism from
1+1 dimensions to 3+1 dimensions.

\subsection{On the possibility of different $K(v)$}

The possibility of modifying the usual Minkowski space-time means,
in the framework of this paper, considering deviations from the special
relativistic formula $K(v)=1-v^{2}/c^{2}$. Since our basic assumption
is that $K(v)$ be a function only of $v=dE/dp,$ we consider deformations
of the type 
\begin{equation}
K(v)={\displaystyle \sum\limits _{k=0}^{n}}a_{2k}\left(\dfrac{v}{c}\right)^{2k},\label{Kmod}
\end{equation}
where $k=0$, $a_{0}=1$ is the usual Galilean term and $k=1,$ $a_{2}=-1$
is the special-relativistic correction. Notice that according to eq.(\ref{Kmod})
$c$ is not necessarily an invariant velocity anymore, being in general
$K(c)\neq0$. However if we restrict our attention to function $K(v)$
such that 
\begin{equation}
{\displaystyle \sum\limits _{k=0}^{n}}a_{2k}=0,
\end{equation}
then $K(c)=0$, so that $c$ is still an invariant speed. If now we
substitute (\ref{Kmod}) in eqs.(\ref{CovSpeedExpress})-(\ref{BoostEnergyAction})
and consider dispersion relations of the type of eq.(\ref{fodr}),
being 
\begin{align}
v & =\dfrac{dE}{dp}\simeq c\left(1-\frac{m^{2}c^{2}}{2p^{2}}-\lambda p\right),\\
\dfrac{d^{2}E}{dp^{2}} & \simeq c\left(\frac{m^{2}c^{2}}{p^{3}}-\lambda\right),
\end{align}
we get in the large momentum limit ( $p\gg mc$ and $m\ll M_{Planck}=\lambda^{-1}c^{-1}$)
\begin{equation}
\left\{ N,p\right\} \simeq\frac{\sum_{k=0}^{n}a_{2k}\left(1-\frac{m^{2}c^{2}}{2p^{2}}-\lambda p\right)^{2k}}{c\left(\frac{m^{2}c^{2}}{p^{3}}-\lambda\right)},\text{ \ \ \ \ \ \ \ \ \ \ }\left\{ N,E\right\} \simeq\frac{\sum_{k=0}^{n}a_{2k}\left(1-\frac{m^{2}c^{2}}{2p^{2}}-\lambda p\right)^{2k+1}}{c\left(\frac{m^{2}c^{2}}{p^{3}}-\lambda\right)}.\label{boostsvgec}
\end{equation}

When $\lambda>0$, that here means subluminal propagation, it is not
possible for $\left\{ N,p\right\} $ and $\left\{ N,E\right\} $ to
vanish. Thus we cannot have in this case neither an invariant momentum/energy
nor an invariant speed of particle, not even for massless particles.
Moreover as soon as the momentum reaches $p\sim\sqrt[3]{m^{2}c^{3}\lambda^{-1}}$
the denominators of eqs.(\ref{boostsvgec}) diverge. There is again
a maximum allowed momentum $p_{\max}\sim\sqrt[3]{m^{2}c^{3}\lambda^{-1}}$
that depends on the particle mass and that can be many orders of magnitude
smaller than the Planck momentum $\left\vert \lambda\right\vert ^{-1}$.
Instead in the case $\lambda<0$ we get that the invariant speed $c$
is reached as soon as the momentum becomes of the order of $p_{I}\sim\sqrt[3]{m^{2}c^{3}\left|\lambda\right|^{-1}/2}$.%
\footnote{Notice that is this case the denominator of $\left\{ N,p\right\} $
does not vanish so that $\left\{ N,p\right\} $ does not diverge.%
} For the same value of the momentum, the numerators of (\ref{boostsvgec})
vanish so that the momentum $p_{I}$ is an invariant momentum and
the corresponding energy an invariant energy. Thus the modifications
introduced in $K(v)$ do not lead to significant differences in the
predictions of the model. Always maintaining the same deformed form
of $K(v)$ it is worth considering different dispersion relations.
If we focus on the lowest orders, in the Planck parameter, of the
dispersion relations of eq.(\ref{MSDDDR}) we get 
\begin{align}
v & =\dfrac{dE}{dp}\simeq c\left(1-\frac{m^{2}c^{2}}{2p^{2}}-\frac{m^{3}c^{4}}{p^{2}}\lambda+\frac{c^{4}m^{2}}{2}\lambda^{2}\right),\\
\dfrac{d^{2}E}{dp^{2}} & \simeq c\left(\frac{m^{2}c^{2}}{p^{3}}+\frac{2m^{3}c^{4}}{p^{3}}\lambda\right),
\end{align}
that suggest that $v=c$ is still the invariant speed for massless
particles, independently on their momenta, and that $p\sim\lambda^{-1}$
remains an invariant momentum (also) for massive particles. Even considering
a dispersion relation that provides the following
\begin{align}
v & =\dfrac{dE}{dp}\simeq c\left(1-\frac{m^{2}c^{2}}{2p^{2}}-\alpha p^{2}m\lambda^{3}\right),\\
\dfrac{d^{2}E}{dp^{2}} & \simeq c\left(\frac{m^{2}c^{2}}{p^{3}}-2\alpha pm\lambda^{3}\right),
\end{align}
where $\alpha$ is a further deformation parameter, we obtain that,
for $\alpha>0$, the invariant speed would be reached at the invariant
momentum $p_{I}\sim\sqrt[4]{mc^{2}/\left|\lambda\right|^{3}}$. Therefore,
in all the analyzed cases, we find only minor departures from the
behavior that the models exhibit with undeformed $K(v)$.

\subsection{On the compatibility of the approach with linear energy addition
rules}

The second point that we want to address is the compatibility of our
approach with the conservation of the energy for composite systems
as analyzed in\ \cite{Mandanici:2007eb}. We have found so far that
the covariance of the dispersion relation implies eq.(\ref{CovRelDisp})
and that the covariance of the evolution equation $v=\left\{ x,E\right\} $,
together with $\left\{ x,p\right\} =1$, implies eq.(\ref{CovSpeedExpress}).
In order to guarantee covariance to the rule of additivity of the energy
(the momentum remaining not additive) according to\ \cite{Mandanici:2007eb}
it is enough to choose 
\begin{equation}
\left\{ N,E\right\} =\pi(p),\hspace{1cm}\text{ }\left\{ N,p\right\} =E\left(\frac{d\pi}{dp}\right)^{-1},\label{laeaction}
\end{equation}
where $\pi(p)$ is a general function of the physical momentum $p$
transforming according the usual special relativistic rules. From
eq.(\ref{laeaction}) (see \cite{Mandanici:2007eb}) follows $E=\sqrt{m^{2}+\pi^{2}}.$
Thus the point reduces to find a dispersion relation compatible with
eqs.(\ref{laeaction}) and with eqs.(\ref{CovRelDisp})-(\ref{CovSpeedExpress}).
In order to achieve this, one has to solve the following system 
\begin{equation}
\begin{cases}
E\left(\frac{d\pi}{dp}\right)^{-1}\frac{d^{2}E}{dp^{2}} & =K\left(\frac{dE}{dp}\right)\\
\frac{dE}{dp}E\left(\frac{d\pi}{dp}\right)^{-1} & =\pi
\end{cases}\label{Sistema}
\end{equation}
always with $E=\sqrt{m^{2}+\pi^{2}}$. Assuming again $K(v)$ to be
undeformed, the second equation of (\ref{Sistema}) is trivially satisfied
for every $\pi(p)$. Instead the first equation implies 
\begin{equation}
\frac{d\pi}{dp}=1\Rightarrow\pi(p)=p.
\end{equation}

Thus the only additive model compatible with our hypotheses, already
in the 1+1-dimensional case, is the usual special-relativistic one.

\subsection{On the possibility to extend the analysis to 3+1 dimensions}

Finally let us came to the possibility to extended our analysis to
3+1 space-time dimensions. Particle speed in a boosted reference frame
transforms according to the formula 
\begin{equation}
v_{i}^{\prime}=v_{i}+\xi_{j}\left\{ N_{j},p_{k}\right\} \dfrac{\partial^{2}E}{\partial p_{i}\partial p_{k}}.
\end{equation}

The request of covariance of the energy-momentum dispersion relation
in 3+1 dimensions becomes 
\begin{equation}
\left\{ N_{i},E\right\} =\dfrac{\partial E}{\partial p_{k}}\left\{ N_{i},p_{k}\right\} .\label{E3d}
\end{equation}

Then the request that the boosted velocity depends only on the particle
speed implies, in the 3+1-dimensional case, the following expression
for the infinitesimal action of the boost on the spatial momentum:
\begin{equation}
\left\{ N_{j},p_{k}\right\} =\left(\mathbf{H}^{-1}\mathbf{K}\right)_{jk},\label{p3d}
\end{equation}
where we have defined 
\begin{equation}
\mathbf{H}_{ik}\mathbf{=}\dfrac{\partial^{2}E}{\partial p_{i}\partial p_{k}},\hspace{1cm}\text{ }\mathbf{K}_{ij}=K_{ij}\left(\frac{\partial E}{\partial p_{i}},\frac{\partial E}{\partial p_{j}}\right).
\end{equation}

For the space-time sector in 3+1 space-time dimensions, we get by
means of the formula 
\begin{equation}
v_{i}=\dfrac{\partial E}{\partial p_{i}}=\dfrac{dx_{i}}{dt},
\end{equation}
and following the procedure of the 1+1-dimensional case, the boost
actions on the space-time coordinates: 
\begin{equation}
\left\{ N_{j},x_{j}\right\} =\delta_{ij}t,\hspace{1cm}\left\{ N_{i},dt\right\} =\left\{ N_{i},\dfrac{dx_{j}}{v_{j}}\right\} =\dfrac{\delta_{ij}-K_{ij}}{v_{j}}dt.
\end{equation}

Again we can maintain the usual Minkowski commutative space-time adopting
the standard 3+1-dimensional expression $\mathbf{H}_{ik}=c^{2}E^{-1}\left(\delta_{ij}-p_{i}p_{j}c^{2}E^{-2}\right)$
and $\mathbf{K}_{ij}=\delta_{ij}-1/c^{2}\partial E/\partial p_{i}\partial E/\partial p_{j}$.
However, in this 3+1-dimensional case, it can be shown by explicit
calculations that the transformation rules constructed using formulas
(\ref{E3d})-(\ref{p3d}) are not compatible with the group structure
of the boost actions in the momentum space.

\section{Final remarks and conclusions}

\label{S10}

Under the hypothesis that the action of the boosts on the speed of
a particle be a function of the particle speed alone, i.e. that $\left\{ N,v\right\} =K(v)$,
we have addressed the issue of the relative rest in doubly special
relativity. Our analysis has shown a way to construct, at least in
1+1 dimensions, double special relativity models in which the relative
rest is not an inertial-observer dependent notion. The key point of
our work is that our results can be obtained without renouncing to
the usual $v=dE/dp$ formula for the particle speed. We have also
argued that, in order our scheme to be fulfilled, there is no need
to renounce to the usual (commutative) Minkowski space-time endowed
with standard energy-independent boosts. From a conservative viewpoint
we have mainly focused on the undeformed function $K(v)=1-v^{2}/c^{2}$.
Within these assumptions we have found that those models in which particle
speed approaches $v\simeq c$ admit invariant momentum $p_{I}$ and
invariant energy $E_{I}$ that depend on the mass of the boosted particle.
For instance, in the case of dispersion relation of the type $E^{2}-c^{2}p^{2}+\lambda p^{3}c^{2}=m^{2}c^{4}$
and for dispersion relation of the type DSR1, the invariant momentum
has the form $p_{I}\sim\sqrt[3]{m^{2}c^{2}\lambda^{-1}/2}$. Considering
the available bounds on photon mass $m_{\gamma}\lesssim10^{-18}eV/c^{2}$
and assuming $\lambda^{-1}\sim10^{19}GeV/c$ we would get for photons
an invariant momentum $p_{I}$ $\lesssim10^{-3}eV/c$, that would be
ruled out by available data. Instead, in the case of the DSR2 dispersion
relation, we find that remains $p_{I}\sim\lambda^{-1}$ independently
on the mass of the particle considered. Maintaining a Minkowski structure
for space-time and a standard form of $K(v)$ also implies that the
transformation laws of angular frequency $\omega$ and wavenumber
$k$ must differ, in general, from those of $E$ and $p$, with influences
on the phenomenology. In fact one would expect no deviations from
special relativity when considering effects such as the Doppler effect
(in space-time coordinates) and/or the time dilatation between inertial
observers. Instead, as discussed, deviations
from special relativity manifest when considering quantities involving
the energy-momentum sector. Different form of $K(v)$ can modify the
phenomenology but, within the class of functions that we have analyzed,
no qualitative changes have emerged with respect to the described
picture. The action of boosts on the energy-momentum space appears
however incompatible with linear addition rules for energies in the
sense of$\,$\cite{Mandanici:2007eb}. We also have found troublesome
to extend the analysis to 3+1-spatial dimensions especially because
of difficulties in maintaining the group structure of the boost actions.

\bibliographystyle{ieeetr}
%\bibliography{Bibliografia3}

\end{document}